\documentstyle[preprint,aps,prc,eqsecnum]{revtex}
\begin{document}
\narrowtext
\draft
\preprint{}
\title{First decay study of the very neutron-rich isotope $^{93}$Br} 
\author{G.~Lhersonneau$^{1,2}$, A.~W\"ohr $^{1,3}$, B.~Pfeiffer$^1$, 
K.-L.~Kratz$^1$, and the ISOLDE Collaboration$^4$\\}
\address{
$^1$ Institut f\"ur Kernchemie, Universit\"at Mainz, D-55099 Mainz, Germany\\ 
$^2$ Department of Physics, University of Jyv\"askyl\"a, P.O.Box.~35, 
FIN-40351 Jyv\"askyl\"a, Finland \\ 
$^3$ Clarendon Laboratory, University of Oxford, Oxford OX1 3PU, U.K.\\
$^4$ CERN, CH-1211 Geneva 23, Switzerland\\}
\date{today}
\maketitle
\begin{abstract}
The decay of the mass-separated, very neutron-rich isotope $^{93}$Br has
been  studied by $\gamma$-spectroscopy. A level scheme of its daughter
$^{93}$Kr  has been constructed. Level energies, $\gamma$-ray branching
ratios and  multipolarities suggest spins and parities which are in
accord with a  smooth systematics of the N = 57 isotones for Z $\leq$
40, suggesting the N = 56 subshell closure still to be effective 
in Kr isotopes.
So far, there is no indication of
a progressive onset of  deformation in neutron-rich Kr isotopes. 
\end{abstract}
\pacs{27.60.+j,23.20.Lv} 
\narrowtext
\section*{Introduction}
\label{sec:level1}
Neutron-rich krypton isotopes are located in an interesting mass region 
where competition of various structures and shapes at low-excitation
energy  occurs. With Z=36, the Kr isotopes lie roughly in the middle
between the  classical  proton shell closures at Z = 28 and 50. It is
therefore expected  that the addition of only a few neutrons to the N=50
shell, would soon lead  to the development of collective features. On
the one hand, this situation,  indeed, corresponds to the Z = 42
molybdenum isotopes which are located  symmetrically to krypton with
respect to the proton midshell at Z = 39. On the  other hand, however,
this picture is at variance with the systematics  established for the
immediate Z$>$36 neighbor isotopes of Kr. Especially,  the even-even
nuclei of strontium and zirconium experience a strong closure  of the
d$_{5/2}$ neutron subshell at N=56. The 2$^+$ states in $^{96}$Zr and 
$^{98}$Zr are quite high with a maximum energy of 1751 keV for $^{96}$Zr 
\cite{NDS96,tblis}. The 2$^+$ states of Sr, although lower than in their 
Zr isotones, are with about 800 keV still fairly high in this region and 
exhibit little collectivity \cite{Machcoll}. Nevertheless, a deformed
minimum  in the potential energy surface at about 1.5 MeV has been
reported for the  N=58 isotones $^{96}$Sr and $^{98}$Zr, based on very
large $\rho^2$(E0)  values \cite{shpcx} and rotational band structures
\cite{Gammasph}. These  deformed structures rapidly come down in energy
with increasing neutron number,  until the N = 60 isotones $^{98}$Sr and
$^{100}$Zr exhibit well developed  rotational ground-state bands
\cite{Gammasph}. Yet, 0$^+$ states at very low energies, at 215 and 331
keV respectively, have been  interpreted as signatures of shape
coexistence \cite{Mach100}. A similar shape  transition is also observed
in Z = 39 yttrium isotopes \cite{L97Yibm,L98Y,RAM99}.  So far, the
heaviest Z = 37 rubidium isotope studied by $\gamma$-spectroscopy  is
$^{94}$Rb, with N = 57 the nearest neighbor isotone of $^{93}$Kr.   Its
low-lying levels were interpreted in the interacting
boson-fermion-fermion  frame as being spherical \cite{LRb94}. 
Published information on levels in Kr isotopes heavier than 
$^{90}$Kr remained rather scarce until recently. It consisted only 
of $\beta$-decay studies leading to levels in $^{91-93}$Kr performed 
by some of us presented in preliminary reports 
\cite{Kratz88b,Pf88,Wo90,Kratz92} and in a PhD thesis \cite{AWphd}. 
However, near completion of this manuscript we became aware of a 
prompt-fission study of the even-even $^{88-94}$Kr \cite{WUpreprint}.  
The data do not show evidence for an increasing softness 
towards deformation, at least until the N = 58 $^{94}$Kr nucleus. 

In addition to $\gamma$-spectroscopy, laser spectroscopic experiments
have been  carried out in this region. The sudden increase of the square
charge radius  $<$r$^{2}$$>$ at N = 60 in the Rb and Sr isotopic chains
was interpreted as  due to the onset of strong ground-state deformation 
\cite{Thibault,Silverans,Mainz2}. In contrast, more recent laser
spectroscopic  measurements on Kr isotopes up to N = 60, did not reveal
such a significant  increase in the $<$r$^{2}$$>$ values \cite{Keim}.
Hence, the onset of  ground-state deformation in the Kr isotopes either
seems to be delayed to  larger neutron numbers, or will occur more
gradually than in the immediate  Z-neighbors. Detailed spectroscopic
studies should be able to answer these  question either, in showing the
existence of deformed excited states coexisting  with spherical states
of similar character as those in the Sr and Zr isotopes,  or in showing
a smooth change from spherical to transitional structures as in  the Mo
isotopes when approaching N = 60. 

For these reasons, spectroscopic investigations of the decays of
neutron-rich  bromine isotopes to their krypton daughter nuclei would be
of great interest.  However, even today the separation of short-lived
halogen isotopes by negative  surface ionization remains an experimental
challenge to ion-source  technology. As one of the last experiments at
the old ISOLDE on-line mass  separator at CERN before the shut-down of
the SC, neutron-rich Kr isotopes  around N = 56 were studied for the
first time and so far only by $\gamma$- and  delayed-neutron
spectroscopy \cite {AWphd}. Here, we report on the decay of  N = 58
$^{93}$Br to $^{93}$Kr, the most neutron-rich isotope of this element 
so far observed by radioactivity. 

\section*{Experiment} 

The $\beta$-decay parent nucleus $^{93}$Br was produced by 600 MeV  
proton-induced fission of uranium. A beam of negatively charged bromine
ions  was obtained by a chemically selective LaB$_6$ surface ion source 
\cite{Vosicki},  and was collected at the detection position on a moving
tape system. The  $\gamma$-rays following $\beta$-decay of $^{93}$Br
were measured with various  detectors: two coaxial Ge-detectors for
transitions up to about 4 MeV, a small  planar Ge-detector for low
energies and a BaF$_2$ scintillator. In addition,  a thin plastic
scintillator was used to detect $\beta$-particles. 

The low threshold of 8 keV of the planar detector allowed the
measurement  of Kr K-X rays necessary for Z-identification as well as
for the application  of the fluorescence method.  In this method,
K-conversion coefficients are  measured by the intensity ratio of the
K-X-ray peak to the one of the  photopeak of the converted transition.
In order to ensure enough selectivity, the method must be applied to
spectra obtained by gating selected transitions.  The coincidence
efficiency was determined by using  a $^{152}$Eu standard source and
on-line decay products including Kr, Rb and  Sr activities in the mass
chains A = 91 to 93. Of particular interest was the  142.4 keV line in
$^{92}$Rb which has  M1 multipolarity \cite{tblis}. 
The low statitics for the A = 93 mass chain limited the use of 
the most accurate timing methods based on coincidences between the 
planar Ge-detector, the plastic-scintillator and the BaF$_2$ crystal.   
Yet, delayed $\gamma$-$\gamma$-t coincidences recorded with the 
large Ge-detectors allowed observation of level lifetimes 
typically longer than 10 ns. 
 
A valid coincidence event was defined by hardware to include  at least
one of the signals from the planar detector or  one of the coaxial
Ge-detectors.   Coincidence events were recorded in list mode with the
GOOSY system  at ISOLDE. They were off-line sorted  on a VAX computer to
create subsets of energy-energy-time  triplets for subsequent gating and
generation of energy or time  projections.

\section{Results}

The identification of the new $\gamma$-lines following the decay of
$^{93}$Br  was made on the basis of the coincidence relations and the
additional requirement that the characteristic X-rays of Kr had to be
present in the coincidence  spectra. In addition to $^{93}$Br and its
decay products, a contamination of  the A = 93 beam due to
[$^{83}$Se$^{10}$B]$^{-}$ molecular ions was observed. This
contamination could be identified unambiguously by comparing the 
intensities of $\gamma$-lines in singles spectra recorded at A = 93 and 
A = 94.  In these spectra, the intensities of the contaminant $^{83}$Se
lines varied  according to the abundance ratio of $^{11}$B/$^{10}$B. The
Z-selectivity of the  negative ion source for bromine was verified by
comparing the $\beta$-delayed  neutron emission probability, P$_n$, of
$^{91}$Br from $\gamma$-spectroscopic  data obtained during this
experiment with earlier $\beta$n-multiscaling measurements \cite{Ewan}. 
The $\gamma$-spectroscopic method is   based on comparison of activities
in the A = 90 and A = 91 chains. The result, P$_n$= 36(2)\%, was found
to be in reasonable agreement  with the value 30(2)\% of ref.
\cite{Ewan}, thus indicating that  there was no sizeable negative
ionization of A = 91 isobars apart  from $^{91}$Br. 
Consequently, it was justified to determine also the P$_n$ value of
$^{93}$Br  by comparing known $\gamma$-intensities of the decays of the
A = 93 daughter  isotopes $^{93}$Kr, $^{93}$Rb and $^{93}$Sr with those
from A = 92 isobars, in particular $^{92}$Rb. In this way, we obtain a
rather large value of P$_n$= 68(7)\% for  $^{93}$Br decay. 
The transitions in $^{93}$Kr populated by 
$^{93}$Br decay are listed in table \ref{tgam} and the decay scheme 
is shown in figure \ref{figscheme}.  

The spin and parity of $^{93}_{35}$Br are not known experimentally. 
In the $_{37}$Rb isotopes ground-state spins vary between I=3/2 and 5/2. 
The spherical shell model predicts odd parity, associated with 
the p$_{3/2}$ and f$_{5/2}$ orbitals, respectively. Both alternatives 
remain for $^{93}$Br. 

According to level systematics of the N=57 isotones, 
the ground state of $^{93}$Kr should have I$^{\pi}$= 1/2$^{+}$. 
This spin was deduced for $^{93}$Kr and $^{95}$Sr from laser spectroscopy 
\cite{Silverans,Keim} and I$^{\pi}$= 1/2$^{+}$ was determined
from transfer reactions for $^{97}$Zr \cite{NDS97}. These spin and
parity correspond to the $\nu$s$_{1/2}$ orbital. 
In the present experiment, the ground-state feeding has been deduced 
from a comparison of $\gamma$-ray intensities of transitions 
assigned to $^{93}$Kr and to its daughters $^{93}$Rb and $^{93}$Sr. 
As for the P$_n$ estimate, this method relies on the assumption that 
the negative-ion source was selective for bromine. With this approach, 
no sizeable $\beta$-branch was found to the ground state of $^{93}$Kr. 
This result is consistent with the assumed change of parity,  
but yet does not preclude positive parity of $^{93}$Br if I$>$3/2.

The level at 117.4 keV is the first excited state in $^{93}$Kr. It is
based on the fact that the $\gamma$-line of this energy  clearly is the
strongest one assigned to $^{93}$Br decay.  A conversion coefficient
$\alpha_K$= 0.058(10) is  measured which implies a dipole transition
($\alpha_K$(E1)= 0.050,  $\alpha_K$(M1)= 0.075, $\alpha_K$(E2)= 0.43).
Shell-model considerations,  which exclude a parity change among
low-spin levels, lead to the  M1 assignment and I$^{\pi}$= (1/2,
3/2)$^{+}$.  

The next levels at 354.8 and 359.4 keV are based on the strong 
$\gamma$-lines at 237.4 and and 242.0 keV seen in coincidence with  the
117.4 keV ground-state transition. The latter level is further 
supported by the observation of a weak 359.4 keV crossover transition 
to the ground state. The conversion coefficients for the above  237.4
and 242.0 keV lines cannot be measured independently of each other. In
the 117 keV gate, both lines contribute to the K-X-ray peak. The
intensity indicates, however, that one of the  transitions is 
predominantly of M1 and
the other one of E2 multipolarity.    
Lifetime measurements indicate that the 237.4 keV line is delayed 
with respect to the 242.0 keV line,  see figure \ref{figtac}.    Since
the feedings of the 354.8 keV and 359.4 keV levels by  $\gamma$-rays are
weak, a  wide gate has been used on one of the channels to include   
the Compton background of unresolved high-energy transitions.  The
threshold was set to 250 keV to avoid triggering by the  lowest-energy
events yielding to a too poor timing response.  
The shift corresponds to a 22(12)~ns halflife for the 354.8 keV  level.
The non-delayed 242.0 keV transition is unlikely E2  but the M1.
Consequently, the  237.4 keV transition is a slightly enhanced E2, the
Weisskopf unit being 29 ns.  Based on these results, the 117.4 keV
level could have a spin  and parity of 3/2$^{+}$, the 354.8 keV level
I$^{\pi}$= 7/2$^{+}$ and the 359.4 keV level  I$^{\pi}$=
(1/2, 3/2, 5/2)$^{+}$.  These spins provide a simple explanation for 
the existence of a crossover transition 
from the 359.4 keV level to the ground state but not from the
354.8 keV level. 
Further levels are placed according to their coincidence relationships
and  energy sum fitting. The highest of the relatively strongly fed
levels at  710.2 keV has a maximum spin of 5/2$^{+}$. However, the
sizeable ground-state  branching rather favors (1/2, 3/2)$^{+}$.  

The vanishing $\beta$-branch to the 1/2$^{+}$ ground-state of $^{93}$Kr
and  the relatively strong feedings of the 354.8 keV (proposed as
7/2$^{+}$) and 359.4 keV levels can be interpreted in a consistent
manner if  a spin and parity of I$^{\pi}$= 5/2$^{-}$ is assumed for
$^{93}$Br.  In this case, a further consequence of the $\beta$-feeding
of  the 359.4 and 709.9 keV levels is that they cannot be 1/2$^+$
states.  We note that the logft values to these levels must be regarded
as lower   limits due to the non-neglectible probability that many weak 
$\beta$-branches to levels between 1.35 MeV and the neutron separation 
energy of S$_n$$\simeq$3.4 MeV have remained unobserved.   In any case,
these logft limits are in agreement with general expectations  for
first-forbidden transitions.   Level properties are shown in table
\ref{tlevels}.

\section{Discussion} 

In order to understand the above results for $^{93}$Kr, it is worthwhile
to  compare them to existing data for neighboring N = 57 isotones. It is
quite  obvious that a dramatic discontinuity in the N = 57 systematics
occurs at  $^{97}$Zr, i.e. at Z = 40 when protons start to occupy the
g$_{9/2}$ subshell  \cite{LMo99}. The level structures of Z$\geq$ 42 nuclei
do not show  signatures of the N = 56 gap associated with the spherical
$\nu$d$_{5/2}$ subshell  closure. In this region, the d$_{5/2}$ and
g$_{7/2}$ neutron orbitals  are remarkably close to each other.
Furthermore, the $\nu$s$_{1/2}$ level  comes down rapidly in energy when
protons are removed and becomes the ground  state in $^{99}_{42}$Mo.
When reaching $^{97}_{40}$Zr  and continuing towards even  lower
Z-values, the first excited state is a 3/2$^{+}$ level while the 
d$_{5/2}$ orbital has not been identified but is not any longer 
a quasiparticle below g$_{7/2}$ \cite{57-59,LZr97}. 
Calculations  performed for the N = 59 isotones $^{97}$Sr and $^{99}$Zr, 
where a similar level  structure exists, indicate a rather complex
character of this 3/2$^{+}$ level,  thus excluding the d$_{3/2}$ single
neutron parentage \cite{LSr97,BrantZr99}.  Its structure rather consists
mainly of g$_{7/2}$ and d$_{3/2}$ neutron  components coupled to core
states. The results from the above theoretical work  may also probably
be applied to the present work on $^{93}$Kr. Accordingly, we propose 
that the 117.4 keV level represents this complex 3/2$^{+}$ state, and
that  the 354.8 keV level is the $\nu$g$_{7/2}$  orbital. 

We tentatively propose assignments for some  higher-lying levels using
systematics of energies and branching ratios  for N = 57 and 59
isotones. These levels are shown in figure  \ref{figsyst}.
The next strongly fed levels, at 359 and 710 keV, could be the  second
3/2$^+$ and 5/2$^{+}$ doublet based on the 2$\otimes$s$_{1/2}$ 
core-plus-particle coupling. According to the IBFM calculations  for
$^{97}$Sr \cite{LSr97}, the 523 keV level is the 3/2$_{2}^{+}$ state. 
Its characteristic feature is a large branching ratio for the 
3/2$_{2}^{+}\to$ 3/2$_{1}^{+}$ transition. By analogy, the 355 keV level 
in $^{93}$Kr and the 1012 keV level in $^{95}$Sr could be the 
3/2$_{2}^{+}$ states in these nuclei.  In $^{97}$Sr the 5/2$_{1}^{+}$
level at 600 keV has an E2  transition to the ground state which is
several times stronger than  the M1 to the 3/2$_{1}^{+}$ level. The 710
keV level in $^{93}$Kr  has similar branching ratios.  In $^{95}$Sr,  
there are two possible 5/2$^{+}$ levels at 681 and 1004 keV
\cite{57-59}. Both $\beta$-decay parents of $^{93}$Kr and $^{95}$Sr  are
5/2$^{-}$ states. Thus, logft values could be used to tentatively favour
the 681 keV level (logft=6.0) over the 1004 keV level  (logft=6.8) as
being the partner level of the 710 keV  level in $^{93}$Kr (logft=5.8). 

From the similarities of the low-lying levels of the N = 57 isotones
of krypton and strontium (see Fig. \ref{figsyst}) we conclude 
that for $^{93}$Kr, as in its neighbor $^{95}$Sr, there is so far no 
signature of increased collectivity at low excitation energy.   
This was already indicated by the rather high energy of 2$^{+}$ states
in even-even  isotopes, 769 keV for $^{92}$Kr$_{56}$ \cite{Kratz88b,Pf88} 
and 665 keV for $^{94}$Kr$_{58}$ \cite{Kratz92,AWphd,WUpreprint}.  
These energies are somewhat lower than in their strontium isotones, 
e.g. 815 keV for $^{96}$Sr$_{58}$, but remain higher than in molybdenum 
isotopes, e.g. 536 keV in $^{100}$Mo$_{58}$.  
We note that a similar 1/2$^{+}$, 3/2$^{+}$ and 7/2$^{+}$ level sequence is
exhibited by $^{97}$Zr, with  however, much larger level spacings due to
the immediate neighborhood  of $^{96}$Zr where the Z = 40 and N = 56
subshell closures  re-inforce each other \cite{LZr97}. In contrast, the
levels in  $^{99}$Mo exhibit a different order, with a 5/2$^+$ first
excited  state associated to the d$_{5/2}$ neutron instead of the 
3/2$^+$ complex level \cite{LMo99}. 
The close similarity of Kr isotopes (Z = 36) with their Sr 
isotones (Z = 38) but the differences with Mo (Z = 42) 
indicates that there is no symmetry with respect to the Z = 39 
proton midshell. This is in contrast to the simple picture of 
describing nuclear structure in terms of the number of valence 
particles or holes only.

One may, however, speculate that -- as observed in the N = 57--59 
Sr and Zr isotopes \cite{57-59,LSr97,BrantZr99} -- also their  
respective Kr isotones will exhibit  
coexistence of spherical states at low energy and levels of 
deformed collective nature at higher energy. 
Spherical-to-deformed shape transitions between N = 58 
and N = 60 were already predicted as early as 1981 by Bucurescu et al. 
\cite{Bucurescu} who had calculated potential-energy surfaces in a 
microscopic-macrosopic approach. Also M\"oller et al. had predicted shape 
coexistence in this mass region within their 1981 finite-range droplet model 
(FRDM) \cite{Mo81,Mopriv}, however not between spherical and 
deformed states but rather between prolate and oblate shapes. More recently, 
similar features were predicted in relativistic mean-field (RMF) calculations 
by Lalazissis et al. \cite{Lalazissis} in an attempt to reproduce the 
laser-spectroscopic measurements of the mean squared charge radii 
($<$r$^2$$>$ of neutron-rich Sr and Kr isotopes 
\cite{Thibault,Silverans,Mainz2}.

The P$_n$ value of $^{93}$Br decay is 68(7)\%. This large
delayed-neutron branch and, in consequence, the weak  $\beta$-feeding to
low-lying states in $^{93}$Kr can easily be understood in  terms of
general nuclear-structure signatures in this mass region. Spherical 
shell-model calculations of Gamow-Teller (GT) decay of $^{93}$Br using
the  quasi-particle random-phase approximation (QRPA) with a
Folded-Yukawa  single-particle potential and a Lipkin-Nogami pairing
model  \cite{Moller-Randrup}  predict the lowest allowed
$\beta$-transitions at 5.2 MeV and 6.0 MeV in  $^{93}$Kr, respectively,
well above the neutron-separation energy of  S$_n$$\simeq$3.4 MeV
\cite{PM,Audi}. These are the  $\nu$g$_{7/2}$$\rightarrow$$\pi$g$_{9/2}$
3QP- and the  $\nu$p$_{3/2}$$\rightarrow$$\pi$p$_{1/2}$ 1QP-transition,
respectively. Hence,  the low-energy part of the $^{93}$Kr spectrum is
only fed by relatively weak  first-forbidden (ff) $\beta$-transitions.
When taking into account the  ff-strength distribution according to the
Gross Theory \cite{Takahashi},  our calculations yield a delayed-neutron
emission branch  of P$_n$= 83\%, in fair agreement with the experimental
observation.  It is interesting to note in this context, that QRPA 
calculations using the deformation parameters of $\epsilon$$\simeq$0.25
as  predicted e.g. by the global mass models FRDM and
ETFSI-1\cite{PM,ETFSI}  already shift some GT-strength below S$_n$ in
$^{93}$Kr, thus resulting  in a P$_n$ values of only about 25\%, in
disagreement with experiment.

\section{Conclusion}

The decay of N = 58 $^{93}$Br has been studied for the first time and a
partial  decay scheme of $^{93}$Kr has been constructed. The low-lying
states extend  the level systematics of the N = 57 isotones. The good
correspondence with the  levels in $^{95}$Sr suggests that the spherical
N = 56 shell gap is still  effective for the Z = 36 Kr isotopes. So far,
no signatures of  an increase of collectivity leading to a transitional
character  have been observed. Shape coexistence in Kr isotopes is
likely to occur for  N$\geq$58 with features similar to those well
established in the Sr and Zr  isotones. The fact that -- according to
laser spectroscopic measurements --  the onset of ground-state
deformation does not occur at N = 60, may imply  that the postulated
deformed levels in the Kr isotopes lie higher  and/or decrease  slower
in energy than in the corresponding Sr and Zr isotones. Hence, it would 
be of the great interest to search for fingerprints of deformation in
even heavier Kr isotopes, although still today this represents a real 
experimental challenge.

\acknowledgments

This work was performed under support of the German BMFT and has been 
completed under support of the Finnish Centre of Excellence Programme 
2000--2005 (Project No.44875, Nuclear and Condensed Matter Physics
Programme). The authors wish to thank Dr.~W.~Urban for communicating 
a manuscript prior to publication.

\newpage

\widetext
\begin{table}
\caption
{List of transitions assigned to the decay of $^{93}$Br 
to $^{93}$Kr. Taking into account the intensity of unplaced transitions, 
100 $\gamma$-intensity units correspond to 60\% branching in this 
decay mode, i.e a 19\% decay branch for $^{93}$Br (P$_n$= 68\%). 
} 
\label{tgam}
\begin{tabular}{rcrrl}          
 Energy &  Intensity & \multicolumn{2}{c}{Placed} & Coincidences \\ 
 $\lbrack$keV] & $\lbrack$\%]    &  from  & to  & \\
\tableline
 117.4 (2) & 100.0 (50)\tablenotemark[1] & 117 & 0 &
             237, 242, 446, 593, 670, (1220)\tablenotemark[3]  \\
 237.4 (2) & 29.6 (25)\tablenotemark[1] & 355 & 117 & 117, 
               (629)\tablenotemark[3]  \\
 242.0 (2) & 59.8 (7.5)\tablenotemark[1] & 359 & 117 & 
        117, 350, 446, 670, 966, 978, 1142\tablenotemark[3]  \\
 349.9 (5) & 3.6 (21)\tablenotemark[2] & 710 & 359 & \\
 359.4 (2) & 3.7 (7)\tablenotemark[1] & 359 & 0 & \\
 446.0 (2) & 6.8 (9)\tablenotemark[1] & 805 & 359 & 117, 242 \\
 592.7 (4) & 10.6 (26)\tablenotemark[2] & 710 & 117 & 117 \\
 669.5 (3) & 3.4 (5)\tablenotemark[1] & 1029 & 359 & (117, 242) \\
 687.9 (2) & 5.3 (4)\tablenotemark[1] & 805 & 117 & (117) \\
 710.2 (2) & 19.7 (18)\tablenotemark[1] & 710 & 0 & \\
 966.4 (7) & 5.9 (39)\tablenotemark[2] & 1326 & 359 & (117, 242) \\
 977.6 (6) & 6.5 (43)\tablenotemark[2] & 1337 & 359 & (117, 242) \\
2103.5 (4) & 19.4 (27)\tablenotemark[1] &  &   &    \tablenotemark[4]  \\
2224.7 (4) & 4.3 (12)\tablenotemark[1] &   &   &    \tablenotemark[4]  \\
3085.8 (7) & 1.4  (4)\tablenotemark[1] &   &   &    \tablenotemark[4]  \\
3606.3 (6) & 7.8 (10)\tablenotemark[1] &   &   &    \tablenotemark[4]  \\ 
\end{tabular} 
\tablenotetext[1] {Intensity from singles spectra. }
\tablenotetext[2] {Intensity from coincidence data only.}
\tablenotetext[3] {Unplaced weak transition with 
 $\delta$E$_{\gamma}\simeq$1~keV and I$_{\gamma}\le$1.} 
\tablenotetext[4] {Unplaced transition assumed a g.s. transition 
 for calculation of $\beta$-feedings.}
\end{table}

\begin{table}
\caption{
Levels in $^{93}$Kr fed in $\beta$-decay of $^{93}$Br. 
Logft values are calculated with T$_{1/2}$($^{93}$Kr)= 102 ms 
{\protect\cite{Kratz88b}}, Q$_{\beta}$= 11.02 MeV
{\protect\cite{Audi}} and P$_n$= 68\% (this work). 
The $\beta$-feedings are for 100 decays to $^{93}$Kr, ie. 
a 32\% decay branch of $^{93}$Br. Unplaced high-energy transitions 
account for a total of about 20\% $\beta$-feeding non listed here. }
\label{tlevels} 
\begin{tabular}{rrrrl} 
 Energy   &  $\beta$-feeding & logft & \quad & I$^{\pi}$ \\
 $\lbrack$keV] &      \% \qquad    &       &       &     \\
\tableline
 0         & $\le$0.5  & $>$7.5 &  & 1/2$^+$  \\
 117.4 (2) &  1.3 (58) &     & & (3/2)$^+$ \\
 354.7 (3) & 18.5 (22) & 5.9 & & (7/2)$^+$ \\
 359.4 (2) & 22.3 (50) & 5.8 & & (3/2, 5/2)$^+$ \tablenotemark[1] \\
 710.2 (2) & 20.7 (35) & 5.8 & & (5/2, 3/2)$^+$ \tablenotemark[1] \\
 805.3 (2) &  7.4 (9)  & 6.2 & & \\
1028.9 (4) &  2.1 (4)  & 6.7 & & \\
1325.8 (7) &  3.6 (23) & 6.4 & & \\
1337.0 (6) &  3.9 (25) & 6.4 & & \\
\end{tabular} 
\tablenotetext[1] {First value is preferred on basis of systematics. }
\end{table}

\newpage
\begin{figure}
\caption
{Partial decay scheme of $^{93}$Br to $^{93}$Kr. In calculating the
$\beta$-feedings it has been taken into account that a fraction of the 
observed $\gamma$-ray intensity is not placed in the scheme, see table 
{\protect\ref{tgam}}. }
\label{figscheme}
\end{figure}

\begin{figure}    
\caption
{Time spectra from $\gamma$-$\gamma$-t delayed coincidences. 
The start signal was given by the selected transitions of 
237.4 keV (closed circles) and 242.0 keV (open circles). 
The stop signal was triggered by the Compton background and photopeaks 
of any coincident transitions of energy above 250 keV. 
The physical time axis is from the right to the left. 
The dashed lines are only a guide to the eye. The absence of a prompt
component in the time spectrum of the 237.4 keV transition assigns 
the delay to originate from the 354.7 keV level and not from another 
level higher in the scheme. }
\label{figtac}
\end{figure}

\begin{figure}  
\caption
{The lowest-lying levels in the N = 57 isotones $^{93}$Kr, 
$^{95}$Sr (top) and the N = 59 isotones $^{97}$Sr and $^{99}$Zr 
(bottom). The similarities suggest the survival of the N = 56 
subshell closure in Kr isotopes, in contrast to the more transitional 
character of Mo isotopes located symmetrically with respect to the 
Z = 29 midshell. } 
\label{figsyst}
\end{figure}

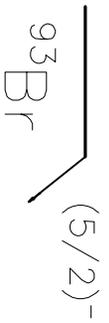
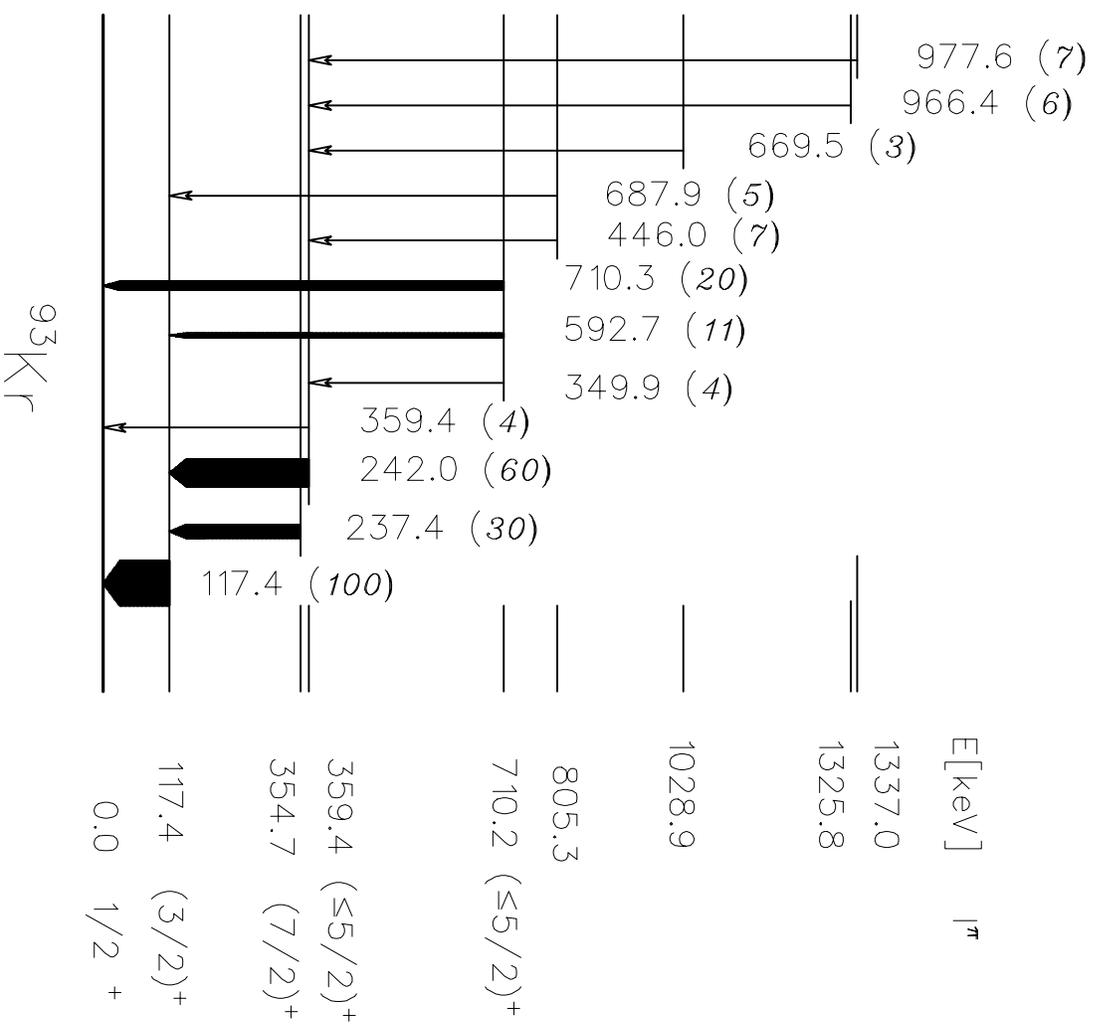
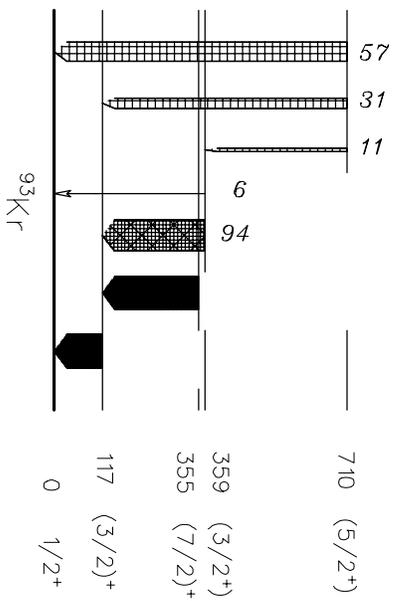
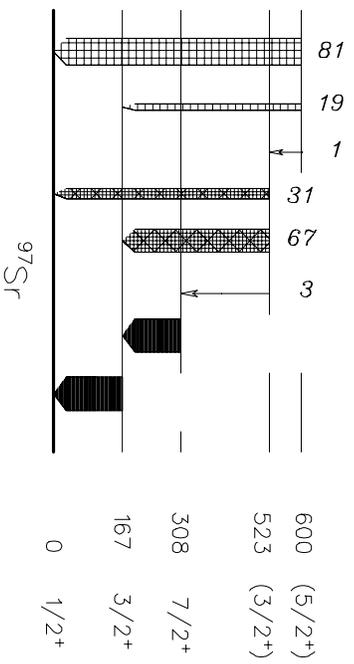
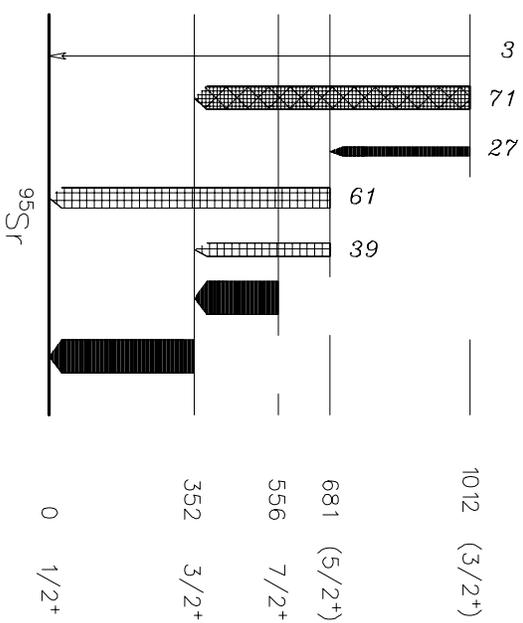
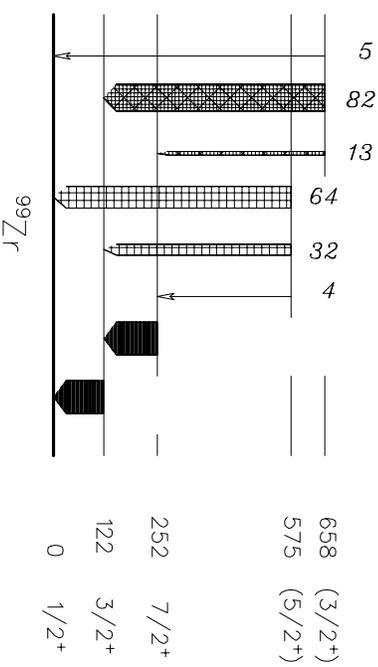
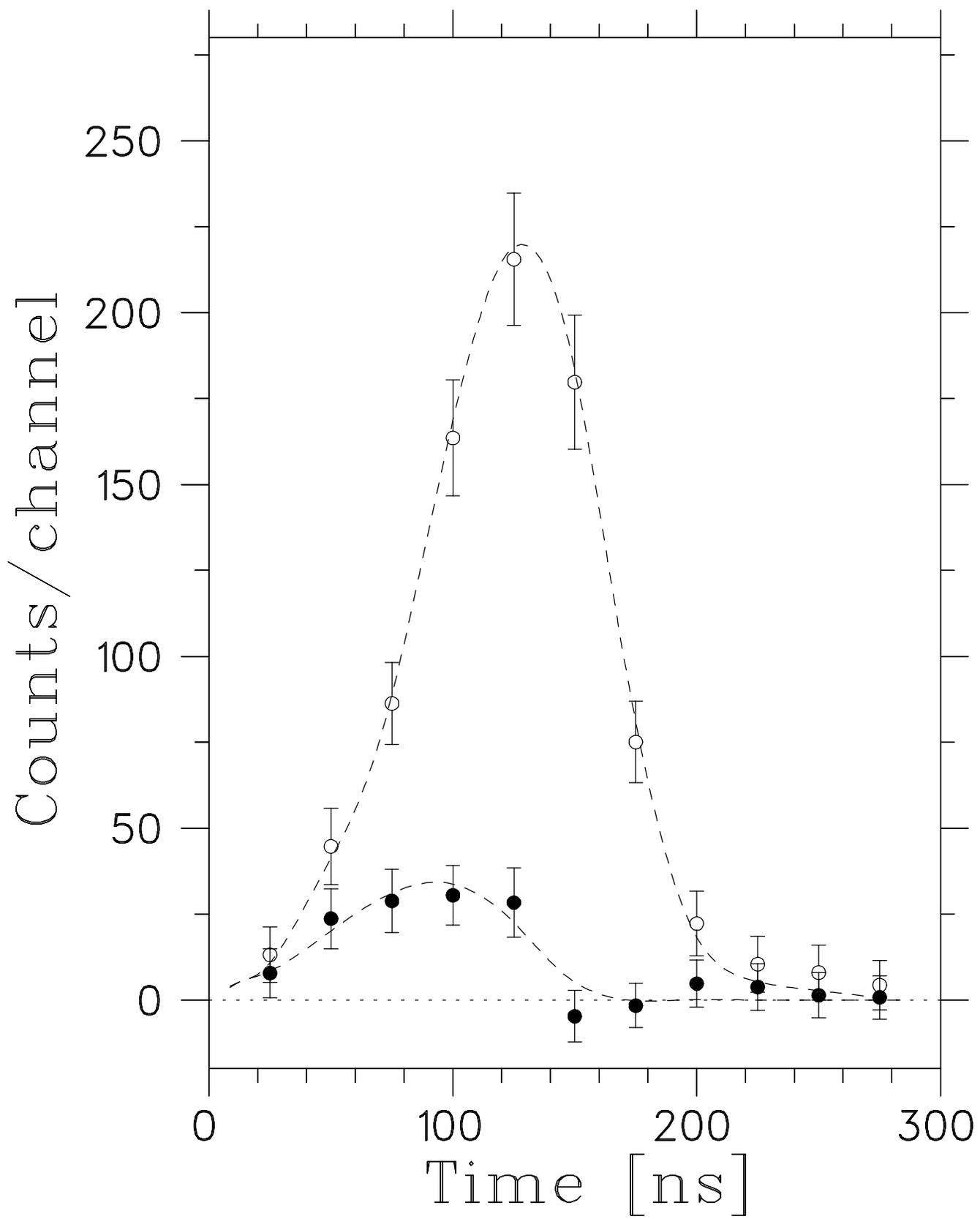
\end{document}